\documentstyle[epsfig]{aipproc}
\begin{document}
\title{Comparative Studies of Line and Continuum Positron Annihilation 
Radiation}
\author{P.A.~Milne $^{1}$, J.D.~Kurfess $^{2}$, R.L.~Kinzer $^{2}$ and 
M.D.~Leising $^{3}$}

\address{
$^{1}$NRC/NRL Resident Research Associate,
Naval Research Lab, 
\linebreak
Code 7650,
Washington DC 20375
\linebreak
$^{2}$Naval Research Lab, Code 7650,
Washington DC 20375
\linebreak
$^{3}$Clemson University, Clemson, SC 29631
}

\maketitle

\vspace{-3mm}
\begin{abstract}
Positron annihilation radiation from the Galaxy has been observed 
by the OSSE, SMM and TGRS instruments. Improved spectral modeling 
of OSSE observations has allowed studies of the distribution of 
both positron annihilation radiation components, the narrow line 
emission at 511 keV and the positronium continuum emission. 
The results derived for each individual annihilation 
component are then compared with each other. These comparisons reveal 
approximate agreement between the distribution of these two emissions. 
In certain regions of the sky (notably in the vicinity of the 
previously reported positive latitude enhancement), the distribution 
of the emissions differ. We discuss these differences and the methods 
currently being employed to understand whether the differences are 
physical or a systematic error in the present analysis.

\end{abstract}
%\begin{center}

\section{Introduction}

Nine years of observations made with the Oriented 
Scintillation Spectrometer Experiment (OSSE) on-board NASA's 
COMPTON observatory (1991-2000)\cite{john93}, eight years of observations 
made with the Gamma-Ray Spectrometer on-board the Solar
Maximum Mission (SMM) (1980-1989)\cite{shar88}, and two years 
of observations made with the Transient Gamma-Ray
Spectrometer (TGRS) on-board the WIND mission (1995-1997) \cite{harr98}
 have been utilized to study the galactic 
distribution of positron annihilation radiation. 
The OSSE instrument featured a 3.8$^{\circ}$ x 11.4$^{\circ}$ FWHM 
FoV, a $\sim$3 x 10$^{-5}$ photons cm$^{-2}$ s$^{-1}$ line sensitivity 
(10$^{6}$ s on-source time), and a 45 keV energy resolution at 511 keV. 
These detector attributes have permitted the first detailed studies of the 
distribution of annihilation radiation in the inner radian of the Galaxy. 
The annihilation of positrons with electrons gives rise to two 
spectral features, a line emission at 511 keV and a positronium 
continuum emission (which increases in intensity with energy 
roughly as a power 
law up to 511 keV and falls abruptly to zero above 511 keV)\cite{ore49}. 
%Studying the spectral signature of positron annihilation radiation 
%requires detectors with superior energy resolution, such as the 
%wide FoV germanium detectors GRIS \& 
%TGRS\cite{gehr91,harr98}. 
The TGRS instrument, which featured a germanium detectors with excellent 
energy resolution, has 
demonstrated that the integrated flux from the inner radian is best 
described as a narrow 511 keV line (FWHM $\leq$ 1.8 keV) and a positronium 
continuum to 511 keV line ratio of $\sim$ 3.6 (which corresponds to a 
positronium fraction of f$_{Ps}$=0.94)\cite{harr98}. 

Purcell et al. (1997) 
(hereafter PURC97)\cite{purc97}
reported results from OSSE/SMM/TGRS studies of the 511 keV line component of 
annihilation radiation. They found the 511 keV emission to be comprised of 
three components; 1) an intense bulge emission, 2) a fainter disk emission, 
and 3) an enhancement of emission at positive latitudes (hereafter called 
a PLE). The PLE was also reported by Cheng et al. (1997)\cite{chen97}, and 
has been interpreted to be an ``annihilation fountain" 
by Dermer \& Skibo\cite{derm97}.
PURC97 characterized the emission via mapping, employing the SVD 
matrix inversion algorithm, and via model fitting, testing the 
combination of a spheroidal Gaussian bulge, a disk that is flat in 
longitude to $\pm$40$^{\circ}$ and Gaussian in latitude (FWHM = 9$^{\circ}$), 
and a spheroidal PLE. The two characterizations differ in the thickness of 
the Gaussian disk (SVD being narrower) and the extension of the PLE. The 
enhancement of the PLE differed between the two characterizations, varying 
from 1.5  x 10$^{-4}$ photons cm$^{-2}$ s$^{-1}$ for the SVD map to 9 x 
10$^{-4}$ photons cm$^{-2}$ s$^{-1}$ for the broad 2D Gaussian PLE (FWHM = 
16.4$^{\circ}$). 
A parallel study of both line and continuum annihilation 
radiation along the galactic plane by Kinzer et al. (1996,2001)\cite{kinz96,kinz01} 
reported that positronium continuum emission is similarly distributed as 
511 keV line emission. The Kinzer studies did not investigate the PLE. 

We report here updates from 
our continuing analysis which extends the study of PURC97 
(see also Milne et al. (1998,1999) \cite{miln98,miln99}). The 
primary differences between current studies and PURC97 are; 1) the inclusion of 
more observations, both archival and data collected after PURC97, and 2) 
reporting maps of the positronium continuum emission 
in addition to the 511 keV line. To extract the positronium continuum 
component from the underlying galactic continuum emission, we 
widened the spectral modeling to include thermal bremsstrahlung and 
exponentially-truncated power-law models. We also removed high-energy 
diffuse continuum emission following a prescription from Kinzer et al. 
(1999), distributing the emission spectrally according to an 
$\alpha$ = -1.65 power-law and spatially according to a 
90$^{\circ}$ x 5$^{\circ}$ 2D Gaussian\cite{kinz99}.
Two maps of the 511 keV line emission and two maps of the 
positronium continuum emission are shown in Figure 1. 

\begin{figure}
\centerline{\epsfig{file=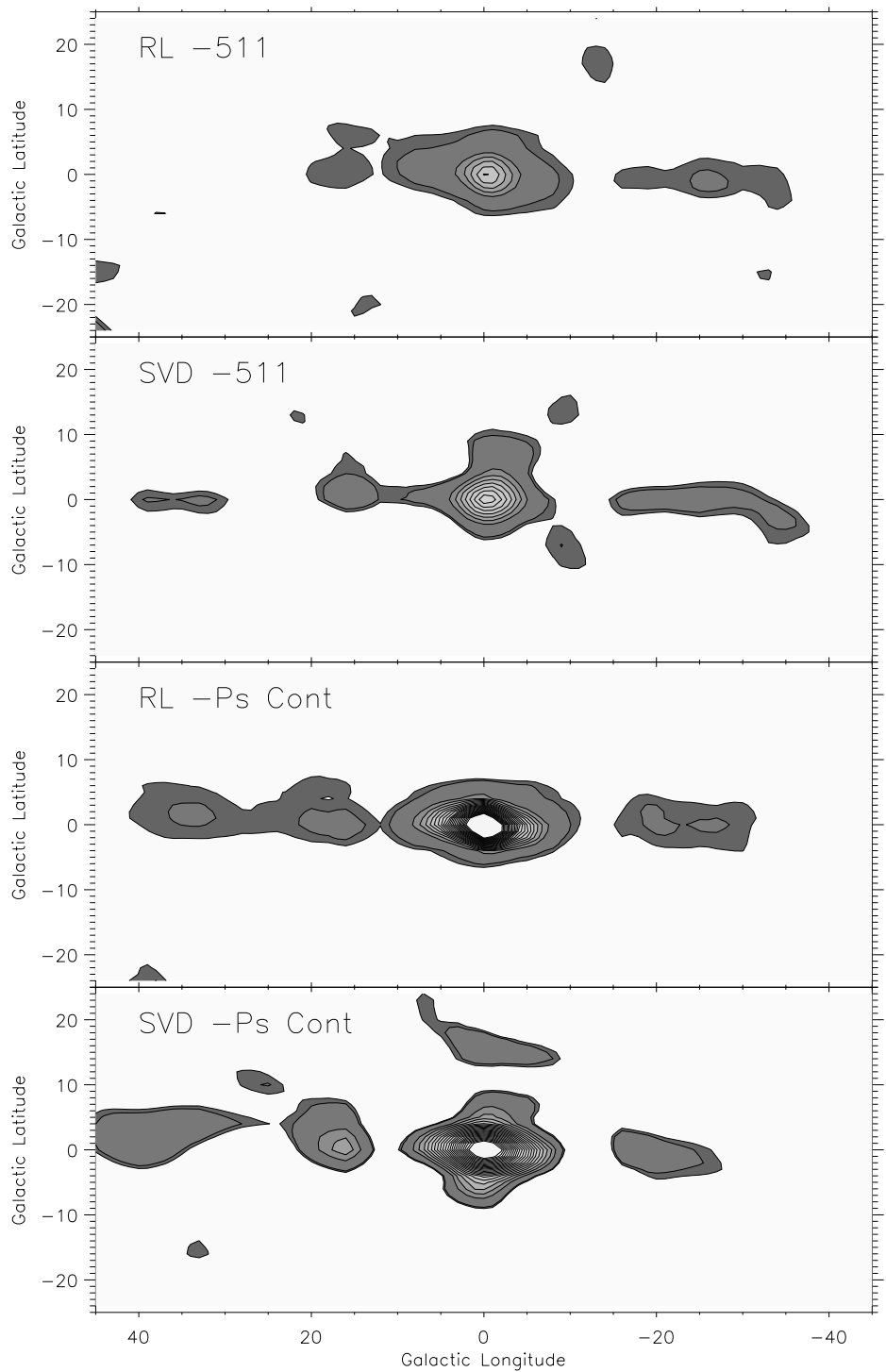, width=13cm}}
\vspace{-.8cm}
\caption{Four characterizations of positron annihilation radiation. The upper 
two panels are Richardson-Lucy and SVD maps of 511 keV line 
emission. The lower two panels are Richardson-Lucy and SVD maps of 
positronium continuum emission. 
}
\end{figure}

\section{Discussion}

Although not identical, the two 511 keV maps share certain 
fundamental features. Both exhibit an intense bulge emission and a 
fainter planar emission. The regions that appear anomalous in the 
RL map (relative to symmetrical bulge and disk emissions) are 
also anomalous in the SVD map. The positronium continuum maps are also 
dominated by an intense bulge and a fainter disk component. Pairings 
of bulge and disk components suggest the same families of solutions. 
Both suggest that the bulge-to-disk ratio can vary from 0.2 -3 depending 
upon whether the bulge component features a halo (which leads to a 
large B/D). The most noticeable difference between the four maps is 
how the PLE is characterized.

Both mapping algorithms suggest enhancement of 511 keV line emission 
from the region reported in PURC97, although at lower flux 
levels ($\leq$ 1.0 x 10$^{-4}$ photons cm$^{-2}$ s$^{-1}$). The broad 
2D Gaussian PLE described in PURC97 is {\it not} an 
acceptable solution; the quality of fit is much worse than solutions 
without a PLE. By contrast, there is no suggestion of a PLE at the 
location reported in PURC97 ($l,b$ = -2$^{\circ}$, 7$^{\circ}$) in either 
of the positronium continuum maps. Shown in Figure 2 are 1$^{\circ}$ 
wide cuts cuts through the four maps, taken at 70$^{\circ}$ relative to the 
negative-longitude galactic plane. It is apparent that the four curves are 
consistent with the same distribution towards negative latitudes, but 
differ at positive latitudes. The lower panel of Figure 2 shows the positive 
latitude cut after subtraction of the mirror negative latitude cut. The 
enhancement at 511 keV has a corresponding {\it deficit} for positronium 
continuum for both mapping algorithms.

An enhancement of 511 keV emission that is due to an additional source of 
positrons (relative to the Galaxy-wide bulge and disk components) would be 
expected to feature an enhancement of both annihilation components. For the 
lowest f$_{Ps}$ value of zero, we would expect the positronium continuum 
emission to be symmetric and the 511 keV line emission to be enhanced.
A positronium continuum {\it deficit} at positive latitudes would not be 
expected. If, alternatively, the PLE is a region with no excess of positrons but 
is where the local f$_{Ps}$ varies from the integrated f$_{Ps}$=0.94 (suggested 
both by OSSE analysis\cite{kinz01}, and 
by wide FoV germanium detectors\cite{harr98}), then the 511 keV line emission could be 
enhanced and the positronium continuum emission could be deficient. However, variations 
of the f$_{Ps}$ {it do not} conserve photon flux. The 1:1 enhancement-to-deficit 
ratio suggested in Figure 2b would not be expected. A third possibility is that the 
enhancement is instead due to the influence of gamma-ray 
sources that corrupt the spectral fitting. Two possible mechanisms for this biasing 
could be from observations made while a source is exhibiting a hard X-ray flare, and/or if 
the source exhibits a previously undetected hard tail. 
The fact that both the line and positronium continuum 
components peak at 511 keV combined with the 45 keV FWHM energy resolution 
of the OSSE instrument mean that 
as many as 30\% of the counts at that energy are due to positronium continuum photons. 
It is unclear whether flaring gamma-ray sources can bias the the spectral fitting 
to a large enough extent to entirely account for the apparent asymmetry in the 
annihilation radiation.

A few compact sources which produce this type of biasing have been identified. At the 
present time, it has not been established which of the three explanations is correct. 
The CGRO/BATSE instrument made observations of these sources which were nearly 
simultaneous with the OSSE observations\cite{harm92}. 
It is an objective of the current analysis effort to determine whether joint 
analysis of OSSE \& BATSE observations of these sources will permit the 
unambiguous extraction of annihilation radiation from this complex environment.
Fortunately, although other regions of the inner galactic radian may be similarly 
biased, the majority of the region is not expected to have been affected. 
 The wealth of information available both from the complete 
OSSE data-set as well as from the expanding data-set of monitoring of compact sources 
with the BATSE instrument may permit definitive statements as to the galactic 
distribution of positron annihilation radiation, particularly of the existence 
of a PLE.

\begin{figure}
\centerline{\epsfig{file=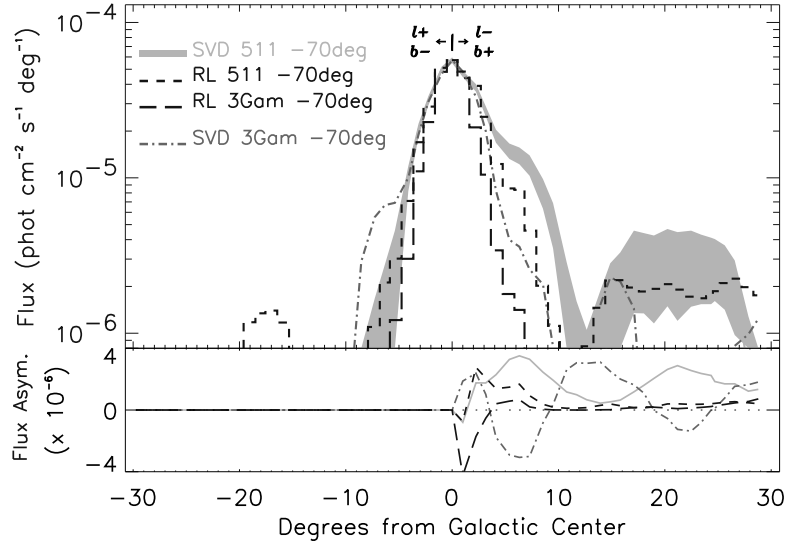, width=10cm}}
\vspace{+.4cm}
\caption{Cuts through the positive latitude enhancement taken 
at 70$^{\circ}$ relative to the negative-longitude galactic 
plane displayed on a logarithmic scale. Shown in 
the upper panel is a 1$^{\circ}$ wide cut through the RL and SVD 
maps of line and continuum annihilation radiation. The SVD -511 
maps shows 1$\sigma$ error bars. The lower panel shows the 
positive latitude portion of the cuts with the mirror negative 
latitude portion subtracted off. 
}
\end{figure}

\end{document}